\documentclass[twocolumn,showpacs,preprintnumbers,amsmath,amssymb]{revtex4-1}

\usepackage{subfiles} 

\usepackage{graphicx}
\usepackage{dcolumn}
\usepackage{bm}
\usepackage{epstopdf}
\usepackage{mathrsfs}
\usepackage{amsmath,amssymb,amsfonts,latexsym}
\usepackage{amsmath,bbold}
\usepackage{color}
\usepackage{array}
\usepackage{slashed}
\usepackage{nicefrac}
\usepackage[colorlinks=true,linktocpage=true,linkcolor=blue,citecolor=blue]{hyperref}
\usepackage[utf8]{inputenc}
\usepackage[mathcal]{euscript}

\def\del{\partial}

\def\sst{\scriptscriptstyle}
\newcommand{\eqn}[1]{Eq.~\eqref{#1}}

\long\def\comment#1{ }

\def\0{{\boldsymbol 0}}
\def\q{{\bm q}}

\def\k{{\boldsymbol k}}

\def\x{{\boldsymbol x}}

\def\and{\qquad\text{and}\qquad}

\def\gmed{g_{\rm med}}

\def\tform{{t_\text{f}}}
\def\tdecoh{t_\text{d}}

\def\med{\text{med}}
\def\jet{\text{jet}}

\def\min{\text{min}}
\def\max{{\text{max}}}

\def\pT{p_{\sst T}}

\newcommand{\beq}{\begin{eqnarray}}
\newcommand{\eeq}{\end{eqnarray}}
\newcommand{\be}{\begin{eqnarray*}}
\newcommand{\ee}{\end{eqnarray*}}
\newcommand{\bal}{\begin{align}}
\newcommand{\eal}{\end{align}}
\newcommand{\rmd}{{\rm d}}
\newcommand{\dd}{{\rm d}}
\newcommand{\rme}{{\rm e}}

\def\rmR{{\rm Re}}

\def\abar{{\rm \bar\alpha}}

\newcommand{\nn}{\nonumber\\ }

\newcommand{\raa}{R_{\rm AA}}

\begin{document}


\title{Cone size dependence of jet suppression in heavy-ion collisions}

\author{Yacine Mehtar-Tani} 
\email{mehtartani@bnl.gov}
\affiliation{RIKEN BNL Research Center and Physics Department, Brookhaven National Laboratory, Upton, NY 11973, USA}
\author{Daniel Pablos}
\email{daniel.pablos@uib.no}
\affiliation{Department of Physics and Technology, University of Bergen, 5007 Bergen, Norway}%
\author{Konrad Tywoniuk}
\email{konrad.tywoniuk@uib.no}
\affiliation{Department of Physics and Technology, University of Bergen, 5007 Bergen, Norway}%

\date{\today}

\begin{abstract}
The strong suppression of high-$\pT$ jets in heavy ion collisions is a result of elastic and inelastic energy loss suffered by the jet multi-prong collection of color charges that are resolved by medium interactions. 
Hence, quenching effects depend on the fluctuations of the jet substructure that are probed by the cone size dependence of the spectrum. 
In this letter, we present the first complete, analytic calculation of the inclusive $R$-dependent jet spectrum in PbPb collisions at LHC energies, including resummation of energy loss effects from hard, vacuum-like emissions occurring in the medium and modeling of soft energy flow and recovery at the jet cone. 
Both the geometry of the collision and the local medium properties, such as the temperature and fluid velocity, are given by a hydrodynamic evolution of the medium, leaving only the coupling constant in the medium as a free parameter. 
The calculation yields a good description of the centrality and $p_T$ dependence of jet suppression for $R=0.4$ together with a mild cone size dependence, which is in agreement with recent experimental results.  
Gauging the theoretical uncertainties, we find that the largest sensitivity resides in the leading logarithmic approximation of the phase space of resolved splittings, which can be improved systematically, while non-perturbative modeling of the soft-gluon sector is of relatively minor importance up to large cone sizes.
\end{abstract}

\maketitle


\paragraph*{Introduction.}
Jets are collimated sprays of energetic particles produced in collider experiments that act as proxies of accelerated quark and gluon degrees of freedom originating from elementary large momentum-transfer processes or decays of massive bosons. In this context, precision computations of QCD jet events play a crucial role in a wide range of fundamental measurements at colliders \cite{Salam:2009jx,Larkoski:2017jix}, including measurements of the Higgs boson properties \cite{Sirunyan:2017dgc} and searches beyond the Standard Model.

In contrast, jet physics in heavy ion collisions probes the discovery frontier to potentially reveal and detail new emergent QCD phenomena in dense partonic systems.
The creation of a short-lived, hot and dense state of deconfined matter, also known as the quark-gluon plasma (QGP), leaves a strong imprint on high-$\pT$ probes \cite{Gyulassy:1990ye,Wang:1991xy}. This phenomenon, commonly referred to as ``jet quenching'', was observed for the first time at RHIC \cite{Adcox:2001jp,Adler:2002xw,Adler:2002tq} and later at the LHC \cite{CMS:2012aa,Aad:2015wga,Aamodt:2010jd,Aad:2014bxa,Adam:2015ewa,Khachatryan:2016jfl}. 
Currently, the exact mechanisms responsible for jet modifications, including details of the energy transport from high-energy to low-energy modes
and color/quantum decoherence of multi-partonic states, are under intense investigation.


 
The basic mechanisms of parton energy loss were understood and formalized in the 90's and implemented for RHIC phenomenology \cite{Zakharov:1996fv,Baier:1996sk,Baier:1996kr,Zakharov:1997uu,Baier:1998kq}, where a number of approximations, in particular for the medium induced radiative spectrum, were then necessary to allow for analytic computations. This introduced a theoretical bias on model calculations, absent from full numerical approaches \cite{CaronHuot:2010bp,Feal:2018sml,Andres:2020vxs}, that could be alleviated 
by incorporating
the two main scattering regimes: the Rutherford scattering regime, dominated by a single hard momentum transfer, and the low momentum regime where multiple scatterings contribute with order one probability. Furthermore, with the measurements
of fully reconstructed jets at the LHC and RHIC, it was soon recognized that higher order corrections accounted for by parton cascades are not negligible. 
The need to address these effects spurred the rapid development of Monte Carlo (MC) event generators \cite{Lokhtin:2005px,Zapp:2008gi,Armesto:2009fj,Renk:2010zx,Young:2011ug,Casalderrey-Solana:2014bpa,He:2015pra,Cao:2017zih,Putschke:2019yrg,Caucal:2019uvr,Ke:2020clc} which, to some extent, rely on modeling of the quantum nature of jet evolution.
In parallel with this computational effort, tremendous conceptual progress has been made in addressing these questions by analyzing the interference structure of two successive splittings within the medium \cite{MehtarTani:2011tz,MehtarTani:2011jw,CasalderreySolana:2011rz,MehtarTani:2012cy,Casalderrey-Solana:2015bww}. 
To leading logarithmic accuracy, it has been shown that the in-medium jet evolution is characterized by an early vacuum parton cascade
whose constituents either get resolved by the medium due to color decoherence, whereas unresolved splittings factorize from the in-medium evolution, losing energy coherently as a single color charge 
\cite{Mehtar-Tani:2017web,Caucal:2018dla}. 
This work aims to address these two challenges within a first-principle analytic framework.

In addition to the modification of the hard components of the jet and their interactions with the plasma constituents, 
there are non-universal contributions to jet observables pertaining to 
how soft jet constituents thermalize in the plasma. 
In analogy with hadronization effects, these non-perturbative contributions are bound to be modeled. This leads us to one of the most important questions in jet quenching physics:
what is the relative magnitude of the uncertainties associated with describing the hard, perturbative structures---that are systematically improvable---and the soft, infrared features of medium-modified jets as a function of their kinematics?
Providing a quantitative answer to this question is crucial if one aims to establish the predictive power of weak coupling techniques in jet quenching phenomenology and probe the transport properties of the QGP. With this work, we aim to provide an answer to this fundamental question.  

In this letter, we report 
a first-principle calculation of the single-inclusive jet spectrum and its cone size dependence in heavy ion collisions where high density effects are resummed to all orders. 
Even though jets with a larger cone do retain a larger fraction of the lost energy, a priori reducing jet suppression compared to a smaller one, we show that resumming the additional energy loss induced by the cone-size dependent jet substructure fluctuations yields a final jet suppression that is very mildly dependent on $R$. The well established connection between energy loss dynamics and coherence effects, which determine the actual resolved phase space of the jet in the medium, allows us to confront our results with high-statistics experimental data merely by constraining the strength of the QCD coupling in the medium. Additional fluctuations on the path and medium density explored by the jet, which vary event by event, are taken care of by embedding our framework into a realistic heavy-ion environment in which the medium is described by the explosion of a liquid droplet of deconfined QCD matter.




\paragraph*{Theoretical formalism.} The spectrum of jets with cone size $R$ in proton-proton collisions is given by the convolution of the initial hard parton spectra with the corresponding fragmentation function. The latter describes the energy
remaining within the jet at different angular resolutions $R$, starting from a large value $R_0\sim 1$. For a steeply falling initial spectrum, it can be written as \cite{Dasgupta:2014yra}
\beq
\label{eq:sigma-pp}
\sigma^{\tiny pp}(\pT,R) =  \sum_{k=q,g} f^{(n-1)}_{\jet/k} (R| \pT,R_0) \, \hat \sigma_k(\pT,R_0)\,,
\eeq
where $n\equiv n_k(\pT,R_0)$ is the power-index of the cross-section of the hard parton with flavor $k$. This is calculated at leading order (LO) at the factorization scale $Q^2_{\rm fac}$, such that $\hat \sigma_k = f_{i/A} \otimes f_{j/A} \otimes \hat \sigma_{ij \to k (l)} $, and involves a convolution of parton distribution functions (PDFs) $f_{i/A}(x,Q_\text{fac}^2)$ with the $2\to2$ QCD scattering cross section $\hat \sigma_{ij \to kl}$.
The moment of the fragmentation function of an initial hard parton with flavor $k$, i.e. $f^{(n)}_{\jet/k} (R| \pT,R_0) = \int_0^1 \dd x \, x^{n} f_{\text{jet}/k}(R|x,R_0)$,
gets both quark and gluon contributions, $f^{(n)}_{\jet/k} = \sum_{i=q,g}f^{(n)}_{i/k}$, due to flavor conversion during the DGLAP evolution \cite{Dasgupta:2014yra,Dasgupta:2016bnd,Kang:2016mcy,Dai:2016hzf}.

Correspondingly, the cross section in nucleus-nucleus collisions (AA) are convolved with a probability distribution $P(\epsilon)$ describing medium-induced energy loss out of the jet cone, and
reads 
\beq
\label{eq:sigma-AA}
\sigma^{\tiny AA}(\pT,R) = \sum_{i=q,g} \int_0^\infty \rmd \epsilon \, P_i(\epsilon) \tilde \sigma^{\tiny pp}_i(\pT+\epsilon,R) \,,
\eeq
where $\tilde \sigma_{i}^{\tiny pp}$ corresponds to the quark/gluon contribution to the total cross section in Eq.~\eqref{eq:sigma-pp} 
(the tilde serves as a reminder that the proton PDFs are replaced by nuclear PDFs).
Finally, the flavor dependent resummed quenching factors (QF) $Q_i(\pT,R)\equiv\int_0^\infty \rmd \epsilon P_i(\epsilon)\tilde \sigma_i^{\tiny pp}(p_T+\epsilon) /\tilde \sigma_i^{\tiny pp}(p_T)$ 
account for the energy loss by a jet with momentum $p_T$ and size $R$ during the passage of a background medium \cite{Baier:2001yt}. In the limit of large power index $n$,
we 
we invoke the asymptotic expansion $\tilde \sigma^{\tiny pp} \propto (\pT+\epsilon)^{-n} \sim p_T^{-n } \rme^{-n\epsilon/p_T}\left(1+{\cal O}(n\epsilon^2/p_T^2)\right) $, which to leading order allows us to identify the QF with the Laplace transform (LT) of the energy loss probability, i.e., $Q(\pT) = \int_0^\infty \rmd \epsilon P(\epsilon) \rme^{-\nu \epsilon } |_{\nu=n/p_T}$, where we have omited the flavor subscript for clarity. 

The factorization \eqref{eq:sigma-AA} reduces trivially to the jet production cross section in the absence of final-state interactions, Eq.~\eqref{eq:sigma-pp}, by setting the quenching factors to unity, $Q_i \to 1$, and replacing the nuclear PDFs by standard proton PDFs. It is justified by the fact that out of cone vacuum evolution takes place at much shorter times than energy loss and was used as a basis for the extraction of the quenching weights from the data \cite{Qiu:2019sfj,He:2018gks}.

A novel ingredient of our setup are the quenching factors $Q_i(\pT,R)$ that resum contributions to the total energy loss of a jet consisting of many color charges that interact with the medium. Every splitting that occurs at short time scales within the medium, gives rise to an additional color current that can scatter with the plasma constituents and source further medium-induced energy loss.
The magnitude of this effect can be gauged by comparing the formation time of a splitting, $\tform = 2/[z(1-z) p_T \theta^2]$, to the characteristic time scale the medium needs to resolve the product of the splitting, namely $\tdecoh = [\hat q_0 \theta^2/12]^{-1/3}$ \cite{MehtarTani:2011tz,MehtarTani:2011jw,CasalderreySolana:2011rz,MehtarTani:2012cy}. Here, $\hat q_0 \equiv \rmd  \langle k_\perp^2\rangle/\rmd t $ is the transport coefficient that encodes medium properties, the so-called jet quenching parameter \cite{Baier:1996sk}. Hence, jet splittings occurring at time scales much shorter than the related medium time scale, that is if $\tform \ll \tdecoh \ll L$, are unaffected by the medium and obey the same properties as vacuum splittings \cite{Mehtar-Tani:2017web,Caucal:2018dla}.
The latter inequality
implies that a splitting with $\theta < \theta_c$, where the critical angle is $\theta_c = (\hat q_0 L^3/12)^{-1/2}$, will not be resolved by the medium. 
 With these considerations in mind, one can show that in the large $n$ limit, owing to the fact that the convolution of energy loss probability distributions reduces to a direct product of quenching factors (in Laplace space), the evolution equation for the resummed quenching weight is \cite{Mehtar-Tani:2017web}
\begin{align}
\label{eq:collimator-eq}
\frac{\del Q_i(p,\theta)}{\del \ln \theta} &= \int_0^1 \dd z \,\frac{\alpha_s(k_\perp)}{2 \pi} p_{ji}^{(k)}(z) \Theta_\text{res}(z,\theta) \nn
&\times \left[Q_j(zp,\theta) Q_k((1-z)p,\theta) - Q_i(p,\theta) \right] \,,
\end{align}
where $k_\perp = z(1-z) p \theta $, $p_{ji}^{(k)}(z)$ are the un-regularized Altarelli-Parisi splitting functions and the phase space constraint is given by  $\Theta_\text{res}(p,R) = \Theta(\tform < \tdecoh < L)$.
Above, it is understood that $p$ is evaluated at $p\equiv p_T$. This distinction is necessary when solving \eqn{eq:collimator-eq} since the initial condition also depends on $p_T$. 
The non-linear evolution equations account for the energy loss of the multi-prong jet substructures that are generated by early collinear splittings. 


The initial conditions for the resummed quenching factors $Q_i(p,R)$ at $R=0$ are the bare quenching factors for single partons.
In this work, we have $Q_i(p,0) = Q^{(0)}_{\text{rad},i} (\pT) Q^{(0)}_{\text{el},i}(\pT) $, where the two bare quenching factors are the LT of the corresponding probability distributions that describe radiative and elastic energy loss \cite{Baier:2001yt,Salgado:2003gb}, For their precise definitions, see Eqs.~\eqref{eq:radeloss} and \eqref{eq:eleloss} below. 
The radiative and elastic energy loss are driven by the transport coefficients $\hat q$ and $\hat e$ \cite{Majumder:2008zg}, respectively, which are related by the fluctuation-dissipation relation $\hat e = \hat q /(4T)$ in  a weakly-coupled plasma 
\cite{Moore:2004tg} (where $\hat e_g = \hat e$ for gluons, and $\hat e_q = \frac{C_F}{N_c} \hat e_g$ for quarks). 
The quenching factor due to radiative energy loss off a single parton is simply the exponential of the LT for a single inclusive gluon radiative spectrum \cite{Baier:2001yt,Salgado:2003gb}.
For our purposes, we should rather consider how single partons contribute to the energy loss of the jet by accounting for the energy that is transported outside of the jet reconstruction cone. 
To this aim, we exploit the wide parametric angular separation between the regime of soft emissions that undergo a rapid turbulent cascade responsible for transporting energy from the jet scale to the medium temperature where dissipation forces take over, and the regime of collimated semi-hard emissions, which experience broadening through collisions with the medium constituents \cite{Blaizot:2014ula,Blaizot:2014rla,Iancu:2015uja,Mehtar-Tani:2018zba,Schlichting:2020lef}. 

The medium-induced gluon radiation spectrum has been computed up to next-to-leading order (NLO) within the improved opacity expansion  (IOE)  in the soft limit \cite{Mehtar-Tani:2019tvy,Mehtar-Tani:2019ygg,Barata:2020sav} and unifies both the BDMPS approach with the GLV/higher-twist formalism \cite{Gyulassy:2000fs,Guo:2000nz}, which has proven to be an important ingredient for phenomenological studies \cite{Feal:2019xfl}. The  IOE was also shown to be very accurate when compared to exact numerical solutions \cite{Andres:2020kfg}. It is expressed as $\dd I_\text{NLO}/\dd \omega = \dd I^{(0)}/\dd \omega + \dd I^{(1)}/ \dd \omega$, with
\begin{align}
\label{eq:dIdomega0}	
&\frac{\dd I^{(0)}}{\dd \omega} = \frac{2\alpha_s C_R}{\pi\omega} \, \ln \left|\cos \Omega L \right| \,, \\
\label{eq:dIdomega1}
&\frac{\dd I^{(1)}}{\dd \omega} = \frac{\alpha_s C_R \hat q_0}{2\pi}  \rmR \int_0^L \dd s\, \frac{-1}{k^2(s)}  \ln \frac{-k^2(s)}{Q^2\, \rme^{-\gamma_E}}  \,,
\end{align}
where $\Omega = (1-i) \sqrt{\hat q /(4 \omega)}$,
$k^2(s) =i \frac{\omega \Omega}{2}[ \cot \Omega s - \tan \Omega(L-s)]$, and the strong coupling constant runs with the typical transverse momentum of the emission, i.e. $\alpha_s = \alpha_s\big((\hat q \omega)^{1/4} \big)$ \footnote{The running coupling is evaluated at leading order with 5 active flavors, and regularized as $\alpha_s(k) = \min[1,2\pi/(\beta_0 \log k/Q_0) ]$, with $\beta_0=23/3$ and $Q_0=0.09$ GeV.}. 
In this expansion, the {\it effective} transport coefficient $\hat q$ differs from the bare $\hat q_0$ by a factor that reflects the full leading logarithmic contribution, i.e.
\beq
\label{eq:qhat-log}
\hat q = \hat q_0 \ln \frac{Q^2}{\mu_\ast^2} \,,
\eeq
where 
$\hat q_0 = \gmed^2 N_c m_D^2 T/(4\pi)$ for a thermal medium in the Hard Thermal Loop (HTL) theory and the lower cut-off scale is $\mu_\ast^2 = m_D^2\, \exp[-2 + 2\gamma_E]/4 $ \cite{Mehtar-Tani:2019ygg,Barata:2020sav}. 
The Debye mass $m_D$ computed at LO in a thermal medium reads 
$m_D^2 = 3\gmed^2T^2/2$ (for three active quark flavors). 
The effective medium scale $Q^2$ depends itself on the amount of rescattering in the medium and can be found by solving the transcendental equation $Q^4 = \hat q_0 \omega \, \ln Q^2/\mu_\ast^2$
\footnote{In order to simplify the numerics, such logarithmic terms are truncated at 1, i.e. here and in Eq.~\eqref{eq:qhat-log} the minimum value of $Q^2$ is $\mu_\ast^2$, throughout this work.}.
In our framework, the medium coupling $\gmed$, the only free parameter that determines energy loss, is to be extracted from the comparison to experimental data.

\begin{figure}
\hspace*{-0.8cm}
\includegraphics[width=0.9\columnwidth]{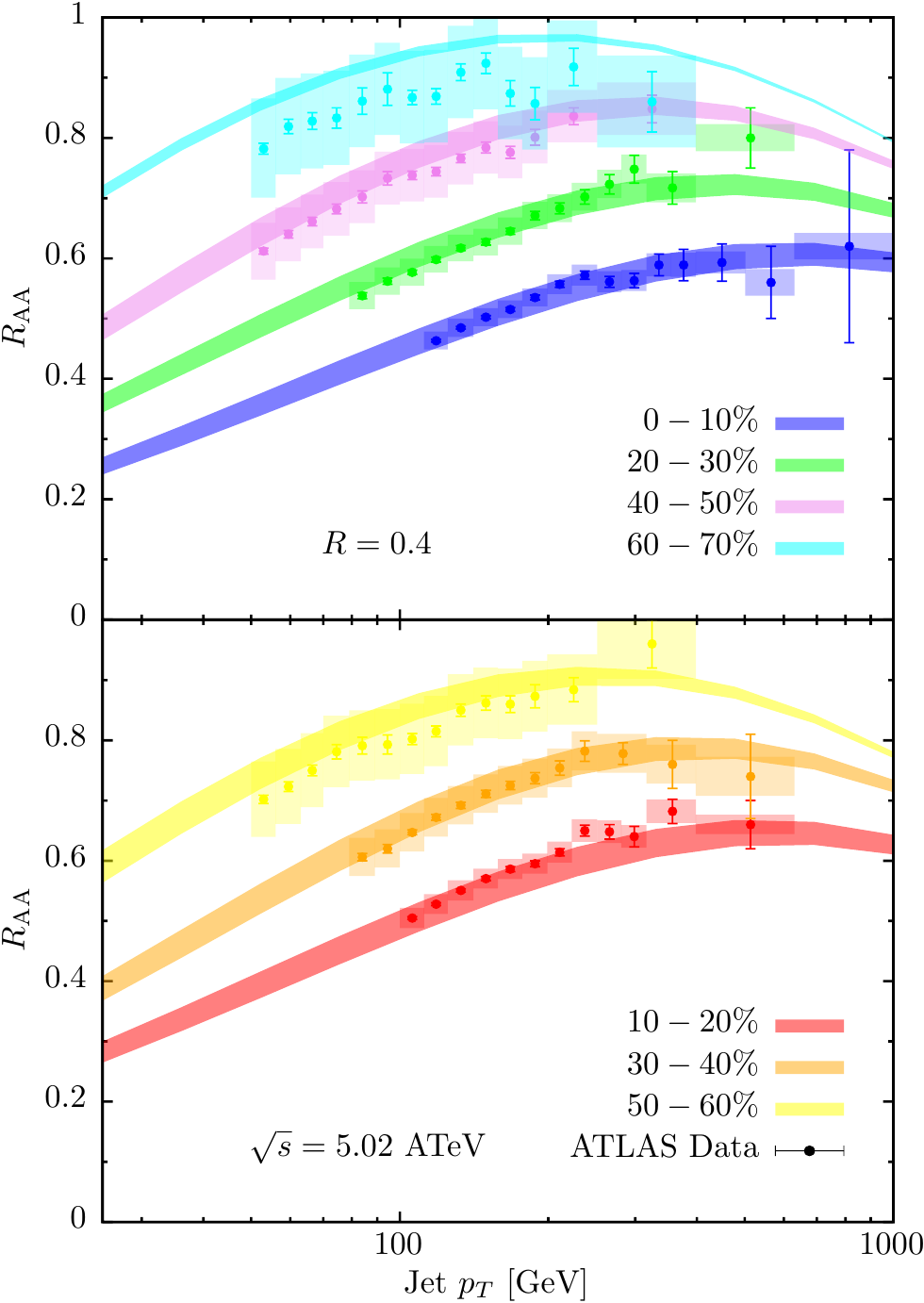}
\caption{\label{fig:jetraa} Calculation of inclusive jet  $\raa$ in PbPb collisions at $\sqrt{s}= 5.02$ ATeV, compared to ATLAS data \cite{Aaboud:2018twu}, for different centralities.}
\end{figure}


We first consider semi-hard gluons that are emitted within the range $\omega_s < \omega \lesssim \omega_c$, where $\omega_c \equiv \hat q_0 \ln(\hat q_0 L / \mu_\ast^2) L^2/2$
corresponds to the maximum accumulated energy through multiple soft scatterings, and $\omega_s \equiv (\gmed^2 N_c/(2\pi)^2)^2 \pi \hat q_0 L^2$ is the energy scale at which emission probability is of order one, determining the onset of turbulent energy loss \cite{Blaizot:2013hx}.
Their broadening distribution reflects the typical transverse momentum kicks received in the plasma.
The fact that the two terms entering the full NLO spectrum are dominated by different kinds of processes has to be reflected in the typical behavior of the respective broadening distribution. In this way, the softer gluons from Eq.~\eqref{eq:dIdomega0}, with $\omega \ll \omega_c$ and small transverse kicks $k_\perp^2\sim \hat q L$, will experience Gaussian broadening,
while the harder emissions from Eq.~\eqref{eq:dIdomega1}, with $\omega \gg \omega_c$ and typically large transverse momenta $k_\perp^2 > \omega/L \gg \hat q L$, where the first inequality arises from demanding that $\tform = \omega/k_\perp^2 < L$, are governed by a power-law behavior, $\sim \hat q_0 L/k_\perp^4$. We assume that the effect of broadening appears as a multiplicative factor 
$B\big(\omega R; Q_{\rm broad}^2 \big) = (\rmd I/\rmd \omega)^{-1} \int_{(\omega R)^2}^\infty \rmd k_\perp^2 \, \rmd I/(\rmd \omega \rmd k_\perp^2)$,
representing the probability for the emitted gluon to be transported out to an angle larger than the jet cone, $\theta>R$, where $Q_{\rm broad}^2$ denotes a characteristic broadening scale. 
This is concretely realized by integrating the broadening probability distribution, derived in \cite{Barata:2020rdn}, for angles larger than the jet cone.
This distribution and the proposed factorized form correctly interpolate between the multiple-scattering and higher-twist regimes.  
The full out-of-cone spectrum consists of two terms, from the IOE expansion up to NLO with their corresponding broadening factors \cite{Blaizot:2014rla}, such that
\begin{align}\label{eq:outofcone-spectrum-final}
\frac{\dd I_>}{\dd \omega} &=  B\big(\omega R; Q_s^2/2 \big) \frac{\dd I^{(0)}}{\dd \omega} \nn
&+  B\big(\omega R; \max \big[Q_s^2,16 \omega/(\pi^2 L) \big] \big)\frac{\dd I^{(1)}}{\dd \omega} \,.
\end{align}
Since emissions can take place anywhere along the in-medium path, one also has to average over the radiation time. This is approximated by simply setting $ Q_{\rm broad}^2 =\hat q L/2$ \cite{Blaizot:2014ula,Blaizot:2014rla} in the first term.
The choice of scale in the second term reproduces the correct behavior of the full GLV spectrum in the high $\omega$ and $k_\perp $ regime up to a logarithmic factor that we neglect. 
In brief, $\rmd I/(\rmd \omega \rmd k_\perp^2)\simeq  \frac{2\bar \alpha \hat q_0 L}{ \pi \, k_\perp^4}  \left(\ldots \right) \approx16 \omega/(\pi^2 L k_\perp^4)\, \times\frac{\dd I^{(1)}}{\dd \omega}$, 
where the ellipses represent the logarithmic contributions and $\dd I^{(1)}/\dd \omega \sim \pi  \abar q_0L^2 / \omega $ is the limiting behavior of Eq.~\eqref{eq:dIdomega1} for $\omega\gg \omega_c$ and $k_\perp\gg \hat q L$. The $B$ distribution was used in the second term to insure proper normalization. For more details see the supplemental material. 

Soft gluons, with $T < \omega < \omega_s$, cascade quasi-instantaneously to the thermal scale \cite{Blaizot:2013hx} and should effectively be treated within hydrodynamics. Their emission rate is therefore not affected by transverse momentum broadening.  Assuming that their distribution becomes approximately uniform in the solid angle around the jet, we account for the possibility that a  fraction of this energy ends up back in the jet cone by modifying $\omega \to \omega (1-(R/R_{\rm rec})^2)$, where the recovery angle $R_{\rm rec}$ is a free parameter \footnote{The fate of the energy and momentum deposited in the flowing plasma is currently under very active investigation. Recent studies show that the distributions of the soft hadrons coming from the wake excited by the jet passage depend on the local background flow \cite{Tachibana:2020mtb,Casalderrey-Solana:2020rsj}; others point to the importance of the interplay between the two wakes of a dijet system \cite{Yan:2017rku,Pablos:2019ngg}. While the implications of these novel observations on our current calculation deserve a deeper study, for the purpose of the present work our ignorance on such non-perturbative effects is encapsulated in the variation of the parameter $R_{\rm rec}$}. An analogous modification is applied to the elastic quenching factor. Emissions at $\omega < T$ belong to the Bethe-Heitler regime, and are not relevant for our present phenomenological application \cite{Wiedemann:2000za,Andres:2020kfg}.

Putting all the pieces together, the final expression for the radiative bare quenching factor reads
\begin{align}
\label{eq:radeloss}
& Q^{(0)}_{\rm rad}(\pT) = \exp \Bigg[ -\int_{\omega_s}^\infty \dd \omega \, \frac{\dd I_>}{\dd \omega} \left(1 - \rme^{-\nu\omega} \right) \nn
& - \int_{T}^{\omega_s} \dd \omega \, \frac{\dd I^{(0)}}{\dd \omega} \left(1 - \rme^{-\nu\omega(1-\left(\frac{R}{R_\text{rec}} \right)^2)} \right) \Bigg] \,,
\end{align}
where the parton flavor index is implicit and $\nu \equiv n/\pT$. We have approximated $\dd I_{\rm NLO}/\dd \omega \simeq \dd I^{(0)}/\dd \omega$ in the soft regime. The bare quenching factor for elastic energy loss is
\beq
\label{eq:eleloss}
Q^{(0)}_{\rm el}(\pT) = \exp\left[- \hat e L \nu \left(1-\left(\frac{R}{R_\text{rec}} \right)^2 \right) \right] \,,
\eeq
also with implicit parton flavor dependence. It results from taking the LT of $\delta(\epsilon-\hat e L(1-R^2/R_\text{rec}^2))$.
We shall see that our results at small cone sizes are not very sensitive to the above modeling of energy recovery.


\paragraph*{Numerical results.}
Using the bare quenching factors for the radiative \eqref{eq:radeloss} and elastic \eqref{eq:eleloss} contributions to energy loss, we numerically solve the coupled evolution equations in Eq.~\eqref{eq:collimator-eq}. The cone-size dependence of the bare quenching factors, through the broadening effects encoded in $\rmd I_>/\rmd \omega$ and resulting in more energy loss for smaller $R$, are to a large extent washed away by the evolution. This is because wider jets have a larger resolved phase space and hence comprise more radiating charges than the narrower jets, effectively hampering energy recovery.

In our numerical computations, we fix the values of the two free parameters of our setup, $\gmed$ and $R_\text{rec}$. The energy recovery parameter $R_{\textrm{rec}}$ has been varied between $R_{\textrm{rec}}=\pi/2$ and $R_{\textrm{rec}}=(5/6) \, \pi/2$, which was estimated from a linearized approach to model the QGP wake for \cite{Casalderrey-Solana:2016jvj,Casalderrey-Solana:2020rsj}.
To constrain $\gmed$ we have compared our results for the widely used nuclear suppression factor for jet production, also known as $R_{\textrm{AA}}$, for jets with $R=0.4$ around $p_T \sim 100$ GeV against high-statistics experimental data from ATLAS for the 0--10\% centrality class of PbPb collisions at $\sqrt{s}=5.02$ ATeV \cite{Aaboud:2018twu}. In order to compute $R_{\textrm{AA}}$, we have taken the ratio between the nuclear and the vacuum spectra, both defined through Eqs.~\eqref{eq:sigma-pp} and~\eqref{eq:sigma-AA}, which comprise a weighted sum of the quark and gluon jet contributions to the full spectrum 
\footnote{In practice, we fit the shape of the initial spectra using the event generator PYTHIA \cite{Sjostrand:2006za,Sjostrand:2007gs}, using EPS09 at LO \cite{Eskola:2009uj} nuclear PDFs for the medium case, where the spectrum at $R=1$ was parameterized as $\dd \hat \sigma^{(k)}/\dd \pT = \sigma^{(k)}_0 \,(p^{(k)}_{{\sst T},0}/\pT)^{n^{(k)}(\pT)}$ and $n^{(k)}(\pT) = \sum_{i=0}^5 c^{(k)}_i \log^i (p^{(k)}_{{\sst T},0}/\pT )$. We then use the LO DGLAP evolution equations to obtain the spectra for R $<$ 1.}. 
Event-by-event in-medium path fluctuations of a jet through the QGP have been taken into account by embedding our framework into a realistic heavy ion environment as simulated in the VISHNU hydrodynamical model \cite{Shen:2014vra}, see the supplemental material for further details. The value of $\gmed$ is thus constrained by the experimental data to be within the range $\gmed \in\lbrace 2.2, \, 2.3 \rbrace$.
We emphasize that the magnitude of quenching is predominantly driven by the emission of copious soft gluons \cite{Baier:2001yt}, see also Tab.~\ref{tab:summary} below.
The extracted parameters yield an average value $\langle \hat q_0 \rangle \simeq 0.41$ GeV$^2$/fm in 0-10\% central PbPb collisions that is well within the perturbative regime, see the supplemental material for more information on the centrality dependence of key parameters.
However, the logarithmic corrections to the bare medium parameters, resulting in $Q^2=14.2$ GeV$^2$ for the factorization scale and $\hat q =2.46$ GeV$^2$/fm, produce a relatively large maximal medium energy scale $\omega_c \approx $ 65 GeV. These effects only become apparent when carefully treating the dominant scattering regimes of the full spectrum, as achieved with Eqs.~\eqref{eq:dIdomega0} and \eqref{eq:dIdomega1}.

\begin{figure}[t!]
\centering
\includegraphics[width=0.9\columnwidth]{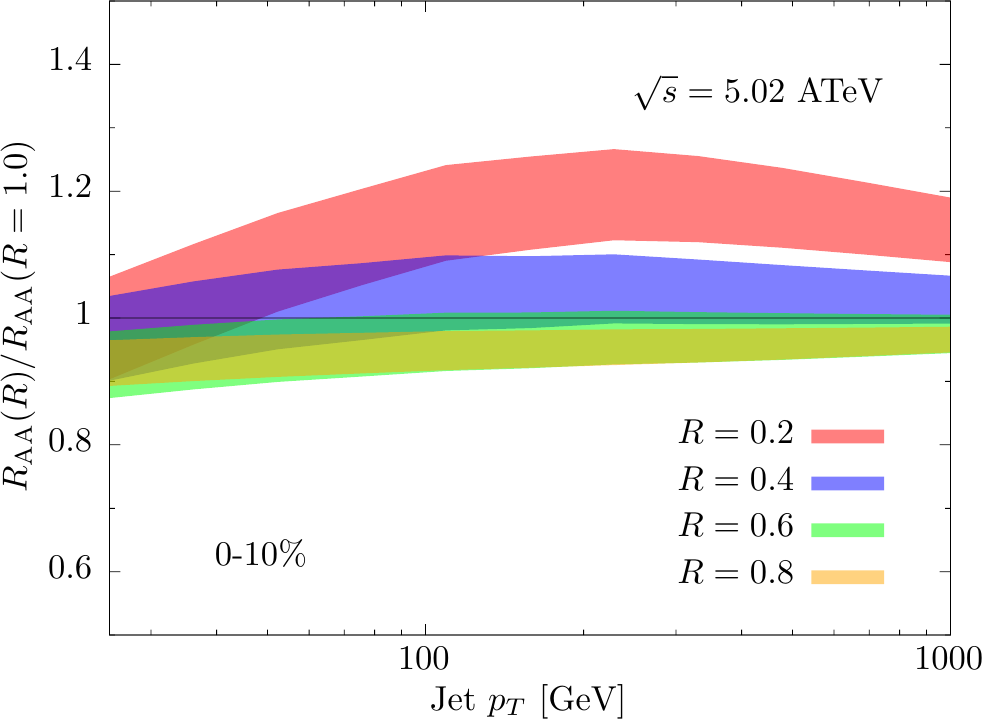}
\caption{\label{fig:doubleratio} Double ratio of inclusive jet $\raa$ for different jet radius $R$ over $\raa$ for $R=1.0$ in PbPb collisions at $\sqrt{s}= 5.02$ ATeV.}  
\end{figure}

We show results for $\raa$ as a function of jet $p_T$ and centrality in Fig.~\ref{fig:jetraa}, confronted against data from ATLAS~\cite{Aaboud:2018twu} for $R=0.4$ jets. We can see that our theoretical results, constrained to describe $\raa$ around $p_T \sim 100$ GeV for 0--10\% only, give an excellent description of both the $p_T$ dependence and centrality evolution of jet suppression. We note that the downturn of $\raa$ at the highest jet $p_T$ is due to the nuclear PDF modifications imprinted in the initial hard parton spectra of PbPb collisions. Finally, we quantify the $R$ (in)dependence of jet suppression 
by taking double ratios of the full results for $\raa$ as in Fig.~\ref{fig:doubleratio}, with the largest size $R=1$ in the denominator. Such notably mild dependence of jet suppression with $R$ is in agreement with ALICE results at low-$\pT$ \cite{Acharya:2019jyg} and with recent experimental preliminary data from CMS at high-$\pT$~\cite{CMS-PAS-HIN-18-014}. 

\begin{table}[t!]
\begin{center}
\begin{tabular}{p{0.2\columnwidth} | p{0.20\columnwidth} | p{0.15\columnwidth} }
Parameter & Variation & Effect  \\
\hline
$\theta_c$ & $[\theta_c/2, \, 2\theta_c]$ & $\lesssim 20\%$   \\
IOE & LO/NLO& $\sim 2\%$  \\
$n$ & $\pm 1$ & $\sim 10\%$   \\
$R_{\rm rec}$ & $[1, \, \infty]$ & $\lesssim 10\%$  \\
$\omega_s$ & $[\omega_s/2,\, 2\omega_s]$ & $\lesssim 8\%$    \\
\end{tabular}
\end{center}
\caption{Summary of the effect of relative change of $R_{AA}$ from varying key parameters for cone sizes $R=0.2-0.6$, see text for further details.}\label{tab:summary}
\end{table}%

\paragraph*{Summary and discussion.} We have provided a first, analytical description of the cone-size dependent jet spectrum in heavy ion collisions at the LHC implemented in a realistic event-by-event setup including nuclear geometry and hydrodynamic expansion of the quark-gluon plasma and accounting for multiple scattering effects. By adequately introducing the notion of single-parton energy loss within the context of a multi-parton object such as a jet, in which the phase space is determined not only by the jet $p_T$ and size $R$ but also on whether splittings are resolved by the medium as determined by color coherence effects, we have shown that the resummation of the quenching factors yield results for jet suppression that are very mildly dependent on $R$. Our procedure to embed our formalism into a realistic heavy ion environment has proven successful, given in particular our good description of the centrality evolution of high-statistics experimental data. 
The various tools and formalism introduced in this work can be systematically improved and will be applied to other jet substructure observables in heavy ion collisions in the future. 

While the error bands provided in the plots above stem mainly from the constraining power of the experimental data at $R=0.4$ on the medium coupling $\gmed$, we would currently like to discuss the sensitivity of our results on the various assumptions made in the setup in order to identify the main sources of uncertainty. Our main findings are summarized in Tab.~\ref{tab:summary} for moderate cone sizes $0.2\leq$ R $\leq0.6$, see also the supplemental material for a full scan of parameters and their centrality dependence.
First, the inclusion of the higher-twist radiative spectrum in the IOE is of mild importance for this observable, since such emissions typically occur at small angles, but it improves the description at high-$\pT$.
Furthermore, as expected, notable bias effects can be identified through the strong sensitivity to the power of the steeply falling spectrum $n$, which point to the importance of higher order terms in the large $n$ expansion that can be calculated systematically.  More importantly, comparing the effect of changing the hard phase space (through $\theta_c$) and the parameters governing the behavior and recovery of soft gluons (through $\omega_s$ and $R_{\rm rec}$), we note that an increased precision in the perturbative sector is still needed before the sensitivity to non-perturbative effects start to dominate. For instance, going beyond leading logarithmic accuracy to compute $\theta_c$ will be important to rigorously study the interesting marked centrality dependence of this critical angle. The importance of the recovery parameter $R_{\rm rec}$ has been gauged between two limiting scenarios of $R_{\rm rec} =1$ corresponding to almost complete energy recovery for large-$R$ jets and $R_{\rm rec} = \infty$ corresponding to no energy recovery. Surprisingly, we find relatively little sensitivity to this parameter at these moderate cone sizes. Conversely, the sensitivity becomes the dominant source of uncertainty only at large-$R$, i.e. $R\approx 1$, jets.

In conclusion, we have demonstrated how the cone-size dependent jet spectrum and $\raa$ is largely governed by the energy loss off hard, resolved jet splittings in the medium through copious, soft gluon radiation and their subsequent broadening out of the jet cone and elastic drag. The results obtained by analyzing the various components of our setup emphasize the importance of analytical tools to guide more sophisticated numerical models, such as MC parton showers.


\vspace{0.5em}
\paragraph*{Acknowledgements.}
Y. M.-T. was supported by the U.S. Department of Energy, Office of Science, Office of Nuclear Physics, under contract No. DE- SC0012704, by Laboratory Directed Research and Development (LDRD) funds from Brookhaven Science Associates and by the RHIC Physics Fellow Program of the RIKEN BNL Research Center. K. T. and D. P. are supported by a Starting Grant from Trond Mohn Foundation (BFS2018REK01) and the University of Bergen. 

\bibliographystyle{apsrev4-1}
\bibliography{microjets}

\newpage

\appendix
\section*{Supplemental Material}

\subsection*{Implementing broadening in the medium-induced spectrum}
\label{sec:broadening-lo}

The broadening distribution is related to the dipole scattering amplitude via a Fourier transform. In a static medium it reads 
\beq\label{eq:broadening-def}
\mathcal P(\k)  = \int \rmd^2 \x  \, \rme^{-i\x\cdot\k -\frac{1}{4}Q_s^2 \x^2 \log \frac{1}{\x^2\mu_\ast^2}} \,,
\eeq
within the leading logarithmic approximation. Here, $\mu_\ast$ represents a model-dependent, infrared scale, see below Eq.~\eqref{eq:qhat-log} for its precise definition in the HTL model of medium interactions. Its asymptotic behavior are as follows
\beq
\mathcal P(\k) \simeq \begin{cases} \frac{4\pi}{Q_s^2}\rme^{-\k^2/Q_s^2} & k_\perp^2 \ll Q_\med^2\\ \frac{4\pi  Q_s^2}{\k^{4}} & k_\perp^2 \gg Q_\med^2 \end{cases} \,.
\eeq
In the latter case, we have neglected additional logarithmic terms. In the regime of multiple scattering, parton splitting take place as a two-stage process where, first, the splitting products separate and decohere and, second, they broaden independently \cite{MehtarTani:2012cy,Blaizot:2012fh}. Therefore, we can easily propose a similar ansatz for the integrated, out-of-cone spectrum
\begin{align}
\omega \frac{\rmd I_>}{\rmd \omega} &= \int_{(\omega R)^2}^\infty \rmd k_\perp^2 \, \omega \frac{\rmd I}{\rmd \omega\, \rmd k_\perp^2} \,, \nn
&\simeq B\big(\omega R;Q_{\rm broad}^2 \big) \times \omega \frac{\rmd I}{\rmd \omega} \,,
\end{align}
where  $B\big(\omega R;Q_{\rm broad}^2 \big) = \frac{Q_{\rm broad}^2}{4\pi} \int_y^\infty \rmd x \, \mathcal P(x)$ with $y=(\omega R)^2/Q_{\rm broad}^2$ and $Q_{\rm broad}$ is the characteristic broadening scale.

The gluon emission regime where $\omega < \omega_c$ is dominated by multiple scatterings. For $\omega > \omega_c$, the dominant contribution to the spectrum is provided by a single, hard scattering with the medium. In this case, the unintegrated GLV ($N=1$) spectrum reads
\begin{align}
\omega \frac{\rmd I}{\rmd \omega \rmd^2 \k}
&= 8 \bar \alpha  \hat q_0 \int \rmd s \int\frac{\rmd^2\q}{(2\pi)^2} \frac{\k\cdot\q}{\k^2(\k-\q)^2 (\q^2+\mu^2)^2}\nn
&\times \left[1-\cos\left(\frac{(\k-\q)^2}{2\omega}s\right)\right] \,,
\end{align}
where $\k=(k_x,k_y)$ is the transverse momentum vector and $k_\perp \equiv |\k|$.
At large energy and transverse momentum, i.e. $k_\perp^2 \gg \omega/L$ and $\omega \gg \mu^2 L$, the spectrum behaves as
\begin{align}\label{eq:glv-highkt}
\omega \frac{\rmd I}{\rmd \omega \, \rmd k_\perp^2} &= \frac{2\bar \alpha \hat q_0 L}{\pi \, k_\perp^4}  \left[\log \frac{k_\perp^2 L}{2\omega} + \log \frac{k_\perp^2}{\mu^2} - 3 + \gamma_E \right] \,.
\end{align}
Given that the integrated spectrum in the limit $\omega \gg \omega_c$ becomes $\omega \frac{\rmd I}{\rmd \omega } = \abar \frac{\pi }{4} \frac{\hat q_0L^2}{2\omega}$,
we can thus express the above unintegrated spectrum as a function of the integrated one
\begin{align}\label{eq:glv-factorization}
 \omega \frac{\rmd I}{\rmd \omega \, \rmd k_\perp^2} &\approx \omega \frac{\rmd I}{\rmd \omega} \times \frac{16 \omega}{\pi^2 L \,k_\perp^4} \big[ \ldots \big] \,,
\end{align}
where the ellipses correspond to the logarithmic corrections in Eq.~\eqref{eq:glv-highkt}. For the purpose of our applications, in the following we will neglect such corrections.

The two discussed limits can be accounted for by expanding the full potential around the harmonic oscillator  \cite{Barata:2020rdn}, where the $\x$-dependent logarithm is neglected in Eq.~\eqref{eq:broadening-def}. The formula for broadening up to next-to-leading order in the improved opacity expansion reads
\begin{align}
\mathcal P&(x) = \frac{4\pi}{Q_s^2} \rme^{-x} \nn
\times &\left\{1 - \lambda \left[\rme^{x} - 2 + (1-x) \big({\rm Ei}(x) - \log(4x) \big) \right] \right\} \,,
\end{align}
where we have defined $x\equiv \k^2/Q_s^2 $, and the expansion parameter is $\lambda = (\ln Q_s^2/\mu_\ast^2)^{-1}$ with $Q_s^2 = \hat q_0 L \ln Q_s^2/\mu_\ast^2$ found implicitly. For our purposes, $\lambda$ is restricted to be within 0 and 1. 
After integrating out the angle, we get
\beq
B(y)= -\lambda + \rme^{-y}\left\{1+\lambda \left[1+y  \big({\rm Ei}(y) - \log 4y \big) 
\right] \right\}
\eeq
with $y \equiv (\omega R)^2/Q_s^2$. In addition, we should also take an average of the production point. A good estimate, at least for the leading order part, is to consider $Q_s^2 \to Q_s^2/2$. The behavior of the $N=1$ spectrum at high $\omega$ is well described by considering instead $\max\big(Q_s^2, 16 \omega/(\pi^2 L) \big)$.


\subsection*{Bare and resummed quenching factors}
\label{sec:quenching-factors}

\begin{figure}
\centering
\includegraphics[width=0.8\columnwidth]{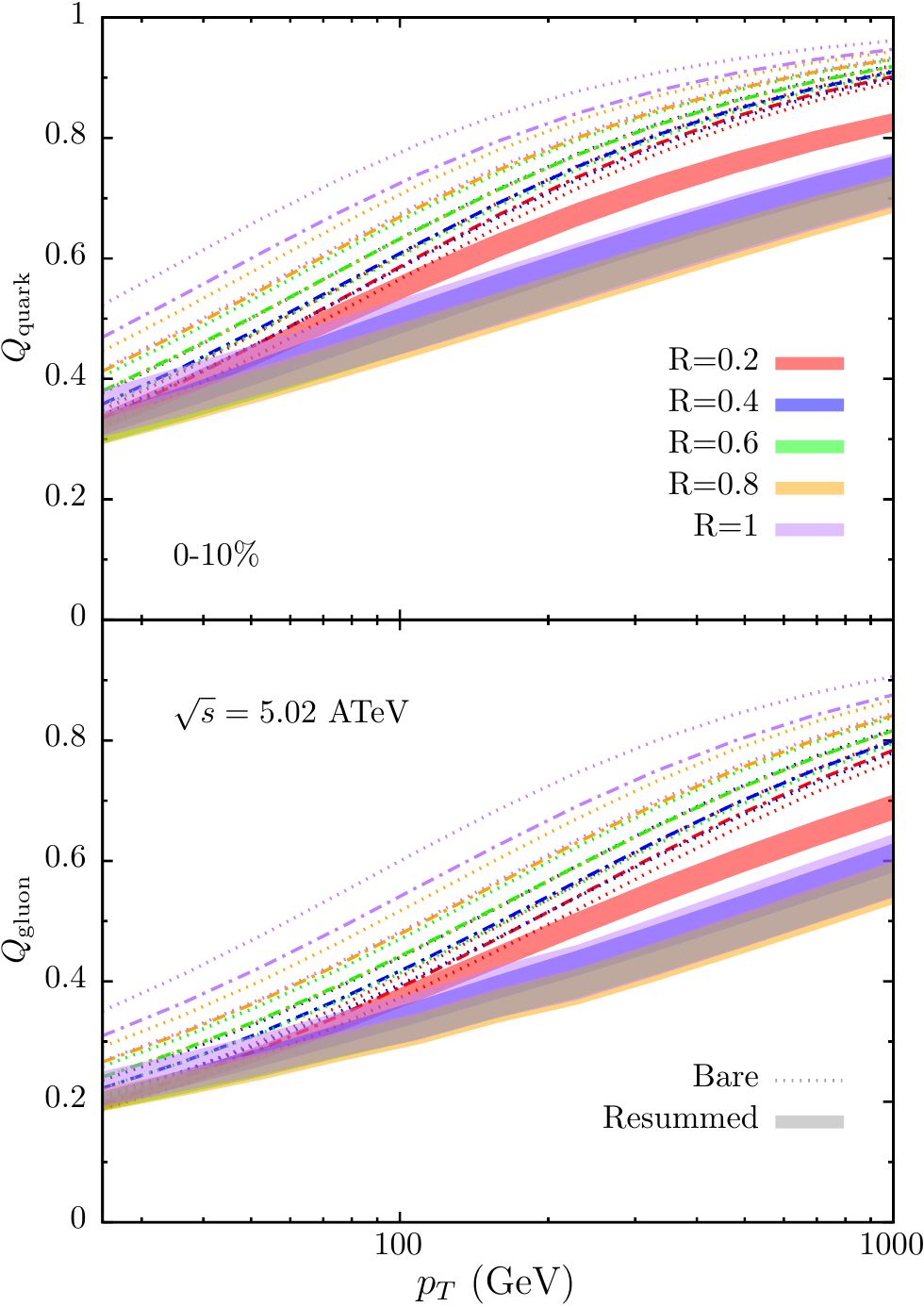}
\caption{Bare and resummed quenching weights for quarks (upper panel) and gluons (lower panel) for central collisions at $\sqrt{s}= 5.02$ ATeV.}
\label{fig:quenching-factors}
\end{figure}

We plot the bare and resummed quenching factors for quarks and gluons in Fig.~\ref{fig:quenching-factors} as a function of $\pT$ for values of the medium parameters corresponding to 0--5\% centrality PbPb collisions at $\sqrt{s}= 5.02$ ATeV. As discussed above, these include both the effect of radiative and elastic energy loss, incorporated in the bare quenching factors as
\beq
Q_{i}^{(0)}(\pT) = Q_{{\rm rad},i}^{(0)}(\pT)Q_{{\rm el},i}^{(0)}(\pT) \,,
\eeq
where $i = q,g$. The $\pT$ dependence appears mainly through the ratio $n_i(\pT)/\pT$, where $n_i(\pT)$ is the power-law indices of the spectra of quark and gluon initiated jets. These weight factors also constitute the initial conditions for the non-linear evolution in Eq.~\eqref{eq:collimator-eq} for the fully resummed weights  $Q_i(\pT,R)$.
The error bands correspond to varying the medium coupling $\gmed\in[2.2, \, 2.3]$ and recovery angle $R_{\rm rec}\in [5 \pi/12, \pi/2]$, as described in the main text.

In Fig.~\ref{fig:quenching-factors} we see that the clear ordering with $R$ of the bare quenching factors (dashed), with more energy loss for smaller $R$, is no longer present in the resummed case (solid).
The energy loss is even enhanced when comparing $R=0.2$ to $R=0.4$. Both for quark initiated jets (top) and gluon initiated jets (bottom), resummed quenching factors show only slight differences among the different $R$.
 
\subsection*{Embedding into a heavy ion environment}
\label{sec:embedding}
\begin{table*}[htb]
\centering
\setlength\extrarowheight{3pt}
\setlength{\tabcolsep}{2pt}
\resizebox{0.75\textwidth}{!}{
\begin{tabular}{c|ccccccccc}
Centrality \,\,& $\hat{q}_0$ [$\textrm{GeV}^2$/fm] & $L$ [fm]  & $\theta_c$ & $w_c$ [GeV] & $w_s$ [GeV] & $\hat{e}$ [GeV/fm] & $m_D$ [GeV] & $T$ [GeV] \\
&&&&&&&&&\\[-1.3em]
\hline
0--5\% \,\, & 0.46 & 5.9  & 0.11 & 80.4 & 21.2 & 1.35 & 0.74 & 0.248\\
&&&&&&&&&\\[-1.3em]
5--10\% \,\,& 0.43 & 5.6  & 0.13 & 63.2 & 17.2 & 1.25 & 0.73 & 0.246 \\
&&&&&&&&&\\[-1.3em]
10--20\% \,\,& 0.41 & 5.0  & 0.15 & 49.3 & 13.9 & 1.17 & 0.72 & 0.242 \\
&&&&&&&&&\\[-1.3em]
20--30\% \,\, & 0.38 & 4.4  & 0.18 & 35.4 & 10.4 & 1.08 & 0.70 & 0.238 \\
&&&&&&&&&\\[-1.3em]
30--40\% \,\,& 0.34 & 3.9  & 0.23 & 23.7 & 7.4 & 0.95 & 0.68 & 0.231 \\
&&&&&&&&&\\[-1.3em]
40--50\% \,\, & 0.29 & 3.3  & 0.28 & 15.7 & 5.2 & 0.82 & 0.65 & 0.22 \\
&&&&&&&&&\\[-1.3em]
50--60\% \,\,& 0.25 & 2.8  & 0.36 & 9.7 & 3.5 & 0.69 & 0.61 & 0.21 \\
&&&&&&&&&\\[-1.3em]
60--70\% \,\,& 0.20 & 2.2  & 0.47 & 5.4 & 2.2 & 0.54 & 0.57 & 0.20 \\[1.ex]
\end{tabular}}
\caption{Average values of the physical quantities entering our calculation. The strength of the QCD coupling constant has been set to $\gmed=2.25$.}  
\label{tab:averagevalues}
\end{table*}

In order to account for the fluctuations in jet energy loss due to the different paths that the jet can explore through the expanding QGP, we need to embed our theoretical framework into a realistic heavy ion background. The necessary steps are the following:
\begin{itemize}
\item Sample the production point in the transverse plane $(x,y)$ using the overlap of the thickness functions of the two nuclei, $T_{AB}(x,y;b)=T_A(x-b/2,y)T_B(x+b/2,y)$, where $b$ is the impact parameter of the nuclear collision and the thickness function is the transverse density of nucleons of the Lorentz contracted nuclei, distributed according to the Woods-Saxon density function (see \cite{Miller:2007ri} for more details on Glauber modeling). 
\item Assign a random orientation in the transverse plane and a random value of rapidity within the range $-2\leq y \leq 2$.
\item Following the path of the jet, compute the integrated values of the necessary physical variables, which in general depend on the local temperature $T$ and fluid velocity $u$, until the jet exits the QGP phase at $T_c=145$ MeV (possible energy loss effects during the hadron gas phase, which could be more relevant for the softer particles \cite{Cassing:2003sb,Werner:2012sv,Bierlich:2018xfw,Dorau:2019ozd}, have been ignored for the moment and will be addressed in the future). The values of $T$ and $u$ are read from event averaged hydrodynamic profiles for the evolution of an expanding droplet of liquid QGP \cite{Shen:2014vra} in PbPb collisions at $\sqrt{s}=5.02$ ATeV for different centrality classes.
\end{itemize}

Given that quantities like the fluid temperature $T$ are given in the local fluid rest frame, we need to consider the actual distance traversed by the jet at each time step within such reference frame \cite{Casalderrey-Solana:2015vaa}:
\beq
dx_F&= \rmd t\, \sqrt{\vec{v}^2 + \gamma_F^2\big(\vec{u}^2 - 2 \vec{u}\cdot \vec{v} + (\vec{u} \cdot \vec{v})^2\big)} \, ,
\eeq
where $\vec{v} \equiv \vec{p}/E$ is the jet axis 
and $\vec{u}$ and $\gamma_F$ are the local fluid velocity and Lorentz factor, respectively. Up to numerical factors, the set of physical variables that we need are found by integrating along the path $\Gamma(t)$ of a jet such as, for example:
\begin{align}
L &= \int_{\Gamma(t)} dx_F \, , \\
\hat{q}_0 &\propto \frac{1}{L} \int_{\Gamma(t)} dx_F\, T^3(x) \, \left( \frac{p\cdot u(x)}{p^0} \right) \, ,
\end{align}
and analogously for the rest of variables: $T$, $m_D$, $\theta_c$, $\hat{e}$ and $w_c$.
Note that, due to the presence of a flowing medium, transport coefficients get a dilution factor $(p\cdot u)/p^0$, where $p$ is the four-momentum of the jet and $u$ the fluid four-velocity \cite{Baier:2006pt}. 

In this way, we obtain a set of representative histories for the jet in-medium path so we can use the event-by-event values of the path integrated physical variables to compute jet energy loss jet-by-jet. Our final results are obtained by averaging over all configurations. 

To provide the reader some guidance on the typical magnitude of these quantities, we show in Table~\ref{tab:averagevalues} the average values for several of the relevant physical variables that enter our calculation, as a function of centrality, where we set $\gmed=2.25$. We point out the interesting correlation between the presence of decoherence effects, for $\theta_c < R$, and complete coherence, $\theta_c \sim R$, as a function of centrality.\\

\begin{table*}[t!]
\centering
\setlength\extrarowheight{3pt}
\setlength{\tabcolsep}{5pt}
\resizebox{0.75\textwidth}{!}{
\begin{tabular}{c|ccccccccccc}
$R$  \,\,& $ n+1 $ &  $n-1$  &  $w_s/2$  & $2 \, w_s$  & $\theta_c/2$  & $2 \, \theta_c$ & $g-0.2$ & $g+0.2$ & $R_{\textrm{rec}}=1$ & $R_{\textrm{rec}}=\infty$ & w/o NLO \\
&&&&&&&&&&&\\[-1.3em]
\hline
$0.2$ \,\,& -0.08  & 0.12  & 0.03  & -0.05  & -0.09  & 0.05  & 0.17  & -0.16  & 0.008  & -0.005  & 0.02  \\
&&&&&&&&&&&\\[-1.3em]
$0.3$  \,\,& -0.08  & 0.11  & 0.05  & -0.06  & -0.12  & 0.11  & 0.19  & -0.17  & 0.02  & -0.01  & 0.02  \\
&&&&&&&&&&&\\[-1.3em]
$0.4$  \,\,& -0.07  & 0.11  & 0.06  & -0.06  & -0.13  & 0.14  & 0.20  & -0.17  & 0.03  & -0.02  & 0.02  \\
&&&&&&&&&&&\\[-1.3em]
$0.5$  \,\,& -0.08  & 0.11  & 0.07  & -0.07  & -0.13  & 0.16  & 0.21  & -0.18  & 0.06  & -0.03  & 0.02  \\
&&&&&&&&&&&\\[-1.3em]
$0.6$  \,\,& -0.08  & 0.11  & 0.08  & -0.07  & -0.13  & 0.17  & 0.22  & -0.18  & 0.10  & -0.05  & 0.02  \\
&&&&&&&&&&&\\[-1.3em]
$0.7$  \,\,& -0.08  & 0.11  & 0.09  & -0.08  & -0.13  & 0.18  & 0.22  & -0.19  & 0.16  & -0.07  & 0.02  \\
&&&&&&&&&&&\\[-1.3em]
$0.8$  \,\,& -0.08  & 0.11  & 0.09  & -0.08  & -0.12  & 0.18  & 0.23  & -0.19  & 0.26  & -0.10  & 0.02 \\
&&&&&&&&&&&\\[-1.3em]
$0.9$  \,\,& -0.08  & 0.11  & 0.09  & -0.08  & -0.11  & 0.17  & 0.23  & -0.19  & 0.49  & -0.13  & 0.02  \\
&&&&&&&&&&&\\[-1.3em]
$1.0$  \,\,& -0.08  & 0.11  & 0.10  & -0.08  & -0.11  & 0.16  & 0.23  & -0.19  & 1.56  & -0.16 & 0.01  \\[1.ex]
\end{tabular}}
\caption{Relative change in $R_{AA}$ around $p_T\simeq 110$ GeV, compared to the central results that use $g=2.25$ and $R_{\textrm{rec}}=\pi /2$, for variations in some of the variables entering our calculation.}  
\label{tab:raa-change}
\end{table*}

\subsection*{Sensitivity of results to systematic uncertainties}
\label{sec:sensitivity}
The calculation presented in this work is based on a perturbative description of hard, vacuum-like and semi-hard, medium-induced emissions. In addition, the regime of soft gluon emissions where thermalization plays an important role are modelled using parametric estimates. In order to systematically quantify the theoretical uncertainties imbued in our modelling, we have performed a scan varying most of the parameters that go into the computation. This is presented as a relative change in $\raa$ at $\pT \simeq 110$ GeV for several jet radius $R$ for the 0-10\% centrality class in PbPb collisions at $\sqrt{s}=5.02$ ATeV, when varying the parameters around the central values $\gmed=2.25$ and $R_{\textrm{rec}} = \pi /2$, as shown in Table~\ref{tab:raa-change}. Clearly, $\raa$ is sensitive to the index $n$ of the hard, steeply falling spectrum. This is a reflection of the bias effect related to jet selection. Next, a large sensitivity is reported in the variation of $\theta_c$ in the phase space of the resummed quenching factors. This points to the importance of pushing beyond the leading logarithmic precision presently implemented. We also report little sensitivity to the precise value of the transition between the semi-hard and soft gluon regimes, encoded in the parameter $\omega_s$. 
Finally, we observe that even though we vary $R_{\rm rec}$ between two extreme values such as $R_{\rm rec}=1$ (full recovery of energy at $R=1$) and $R_{\rm rec}=\infty$ (no recovery at all, for any $R$), these non-perturbative effects remain moderate ($\lesssim 10$\%) up to relatively large cones of around $R=0.6$.

\end{document}